# Bipedal nanowalker by pure physical mechanisms

Juan Cheng[1#], Sarangapani Sreelatha[1#], Ruizheng Hou[1,3], Artem Efremov[1], Ruchuan Liu[1,2,4], Johan RC van der Maarel[1], Zhisong Wang[1,2,3*]

[1]Department of Physics, [2]NUS Graduate School for Integrative Sciences and Engineering, [3]Center for Computational Science and Engineering, National University of Singapore, Singapore 117542

[4]Mechanobiology Institute, National University of Singapore, Singapore 117411

*Corresponding author (phywangz@nus.edu.sg)

# ABSTRACT

Artificial nanowalkers are inspired by biomolecular counterparts from living cells, but remain far from comparable to the latter in design principles. The walkers reported to date mostly rely on chemical mechanisms to gain a direction; they all produce chemical wastes. Here we report a light-powered DNA bipedal walker based on a design principle derived from cellular walkers. The walker has two identical feet and the track has equal binding sites; yet the walker gains a direction by pure physical mechanisms that autonomously amplify an intra-site asymmetry into a ratchet effect. The nanowalker is free of any chemical waste. It has a distinct thermodynamic feature that it possesses the same equilibrium before and after operation, but generates a truly non-equilibrium distribution during operation. The demonstrated design principle exploits mechanical effects and is adaptable for use in other nanomachines.

A dozen artificial track-walking nanomotors[1-12] have been reported since 2004, largely inspired by biomolecular walkers[13] in living cells. Compared to molecular shuttles and rotors[14] in a localized setup, nanowalkers produce long-range directional motion on open tracks. This capability has led to technological applications like nanoscale assembly lines[15], walker-guided surface patterning[16], and walker-mediated organic synthesis[17]. Despite these successes, the design principles by which these artificial walkers rectify a directional motion remain primitive compared to the mechanistic wealth found in natural walkers[13, 18]. With rare exception[7, 10], the artificial walkers change the chemical identity of the track (e.g. burn-the-bridge methods[4-6, 8, 11, 12]) or the environment (e.g. addition/removal of multiple species[1, 2]). These walkers carve a landscape (i.e. track and environment) into a downhill path that switches from one chemical composition (and associated thermodynamic equilibrium) to another to steer a direction. Furthermore, all reported walkers produce chemical wastes, which limits biomedical applications. It is desirable for a nanowalker to navigate a landscape without changing its identity and without producing chemical wastes. The direction instead must be induced by driving the walker-track-environment system away from the single equilibrium via a pure physical action. Here we demonstrate such a nanowalker using a design principle[19] derived from natural walkers[20, 21]. This nanowalker, powered by light, also extends the study of remotely controlled micro/nanomachines[12, 14, 22, 23].

The walker and track are made of six single-stranded DNA molecules as shown by Fig. 1 A-C. The walker consists of two strands (MS1, MS2) that hybridize to form a ~ 6.8 nm long rigid duplex (D3-D3*). Two identical single-stranded legs (5-bases long D1 plus 20-bases long D2) are connected to two ends of the duplex via a 4-bases long single-stranded linker (S1). The track is made of another two DNA strands (TS1, TS2) that each contain a foothold sequence (D1* or D2*) and a hybridization sequence (D4*-D5* or D4-D5). Multiple TS1-TS2 pairs hybridize into periodic double-stranded duplexes D4-D4* (~ 25.5 nm) and D5-D5* (~ 5.1 nm), which form the main body of the track. Repeated pairs of D1* and D2* overhangs between these duplex parts form the footholds. Three extra strands (ES1, ES2, ES3) were introduced to stabilize the track at both ends.

A leg of the walker binds to the track by hybridizing with either the D1* or D2* foothold. The ensuing duplexes D1-D1* and D2-D2* are ~ 2 nm and ~ 6.8 nm long, respectively. Each pair of the D1* and D2* footholds, sandwiching the D5-D5* duplex part, may be regarded as a composite binding site, since it is the only domain of the track capable of forming thermodynamically stable duplexes with the walker's legs. Drawing from foothold D1* to D2* within a composite site points a unique end of the track (called the plus end).

The walker can be operated by any means that breaks the D2-D2* duplex without destabilizing the D1-D1 duplex. Here we develop a light-powered version in which the leg's nucleotide backbone contains nine light-responsive azobenzene moieties in the D2 segment but none in D1. The operation is achieved by alternate irradiation of UV and visible light: UV-light absorption by the azo-moieties creates a high-energy *cis* form that dissociates the D2-D2* duplex; visible light absorption switches the moieties back to the ground-state *trans* form that maintains a stable D2-D2* duplex. The nucleotide sequence of azo-carrying D2 segment is taken from a previous study by Asanuma et al.[24], in which a reversible duplex dissociation/formation was demonstrated. The sequences for other DNA strands are listed in Online Supplement.

The walker-track was assembled from individual strands using a stepwise procedure (Supplement). Intermediate complexes were analyzed using native PAGE (polyacrylamide gel electrophoresis), and the final products were purified using Qiagen gel extraction kit. To assemble tracks with the quencher at the plus end, the track strands (TS1, TS2) were first mixed and annealed; then the end strands (ES1, ES3 and the quencher-carrying ES2) were added to terminate track growth. The complete tracks of a certain length were finally purified from one of multiple bands of the unpurified samples by the right molecular weight to remove other strands/complexes. The presence of the quencher in thus purified tracks was confirmed by a fluorescence drop when the purified tracks were mixed with dye-carrying walkers, as shown by the different fluorescence levels in Fig. 3C and in Fig. 3A, B.

Five products were identified and purified, including the walker, two track species carrying two or three composite sites, and two walker-track binding complexes involving either track species. The five purified complexes each appeared as a single band with expected mobility in the gel image (Fig. 2B), which confirms their formation and thermodynamic stability.

Direction of the walker originates from the use of a small size for the walker so as to restrict its access to inter-site bridge states in which the walker's two legs are bound to two adjacent composite sites. These bridge states are necessary intermediates for the walker's inter-site steps, and four such states are possible (illustrated by B1– B4 in Fig. 1D). The B3 and B4 states are not accessible because the walker is too short to make the respective bindings. For B3, the walker's single-stranded segments have a total contour length (maximum stretch) of $l_C \approx 6$ nm but the extension required to form the state is $l \approx 10$ nm. For B4, the maximum stretch is also short of the required extension ($l_C \approx 20$ nm $< l \approx 22$ nm). However, the walker is long enough for bridges B1 and B2 ($l_C \approx 20$ nm $> l \approx 12$ nm for B1; $l_C \approx 34$ nm $> l \approx 24$ nm for B2). The overall free energy for the state B1 is lower than that for B2, since the leg-track hybridizations are two D1-D1* duplexes for B2 but a D1-D1* plus a longer D2-D2* for B1. Besides the bridge states, a loop state is possible in which the walker's two legs are hybridized with the D1* and D2* footholds within a composite site (B0). The free energy of the loop is below that of B1 because both states have the same leg hybridizations but the walker's single-stranded segments are less stretched in the loop. The state hierarchy from the length analysis is compatible with a quantitative mechanical model (Supplement).

The walker gains a direction towards the track's plus end under the light operation, as shown by panels 4-8 of Fig.1E. By Boltzmann's law, the low-energy B1 state predominates the walker's inter-site bindings on the track prior to operation. This state is asymmetric – the leg to the plus end (referred to as the leading leg hereafter) is in D1-D1* duplex and the trailing leg in D2-D2* (panel 4). A UV irradiation has a chance to dehybridize the trailing leg off the track but not the leading leg though both legs are chemically identical. This selective rear leg dissociation produces a single-leg binding state, in which the D2 segment of the track-bound leg may form close contacts with the D2* foothold at the same composite site (panel 6). When a subsequent irradiation of visible light restores the leg's hybridization capability, D2-D2* hybridization readily occurs to drive the leg's migration from the D1* to D2* foothold (panel 7). After this plus-end-directed migration, the dissociated leg either hybridizes with the forward site to resume the bridge B1 (panel 8) or hybridizes with the nearby D1* to form the intra-site loop B0. In either case, the walker moves towards the plus end. The direction is not compromised by the stochastic nature of light absorption and the ensuing molecular processes, because the selective rear leg dissociation is essentially a ratchet effect[25, 26]. The direction is neither compromised by occasional occurrence of the loop and the bridge B2: the former is readily converted to B1 by the light operation (panels 1-4); the latter is symmetric, irresponsive to the operation, and spontaneously decays to B1 via leg migration (Fig. 1D).

The walker's motion was detected using a fluorescence method (Supplement). The walker's legs were each labeled with a fluorescent dye (FAM); the track was labeled with a quencher at the plus end so as to reduce the dye's emission upon the motor's arrival at the end (Fig. 1). Two types of fluorescence experiments were done with an equimolar mix of walker and purified track containing either two or three composite sites. Before each experiment, the walker-track sample was incubated for 24 hours to ensure thermal equilibrium in walker-track binding. The fluorescence of the incubated sample was monitored over a period of time; the flat signal confirms the equilibrated binding (Fig. S1). The equilibrium fluorescence was used to benchmark each experiment (i.e. data for time zero in Figs. 3, 4). The fluorescence intensity was measured at 25° C for submicromolar concentrations of walker/tracks.

When multiple rounds of alternating UV-visible irradiation were applied to the walker-track sample, the fluorescence drops successively in both the two-site and the three-site experiments (Fig. 3). To test influence of photobleaching that reduces the fluorescence as well, the same light operation was applied to the same amount of walker sample carrying the dye but without the track carrying the quencher. This control experiment yielded a constant fluorescence showing a negligible photobleaching. Thus the fluorescence drop reliably indicates a plus-end accumulation of the walker under the light operation.

The three-site experiment and two-site experiment were both repeated three times using a newly prepared walker-track sample for each repeat. The pattern of successive fluorescence drop recurred for all repeats (Fig. S2). For a quantitative analysis, the fluorescence spectra were integrated over wavelength and then averaged over the repeats. The equilibrium fluorescence signal thus obtained is (51.1±0.8)% or (68.1±0.5)%, for the two-site and three-site tracks respectively, of the signal from the same amount of walkers in the absence of tracks (Fig. S2). Under irradiation operation, the three-site signal further drops ~ 12.3% from the equilibrium fluorescence over 3-hours operation (Fig. S2) and the two-site signal drops ~ 54.4% over 10-hour operation (Fig. 4). The equilibrium signals were analyzed using the Boltzmann distribution (Supplement). A small free-energy gap of ~ 4.95 kJ/mol between the bridge (B1) and the loop (B0) was deduced from the data. This quantitative analysis found the same energy order for the two states as the previous length argument.

The equilibrium fluorescence from the two-site and three-site experiments is very close to 1/2 and 2/3 of the total emission in absence of the tracks. The two ratios indicate around one half or one third of all the walker legs being bound to the plus-end site where the quencher is tethered. The other sites on the two-site or three-site track, with identical footholds, bind comparable percentages of legs, if not more at the middle site. Thus, almost 100% of the walker legs available in the equilibrated samples are track-bound, consistent with the single bands observed for the walker-track complexes (Fig. 2, lanes 4 and 5). The > 50% fluorescence drop of the two-site experiment indicates a plus-end accumulation of > 75% of the walker legs. Such above-equilibrium accumulation must be at the expense of population at other sites of a track. Thus the light operation drives a walker-track sample away from equilibrium in an asymmetric way. The observed fluorescence drop is recoverable after the walker stops operation. Fig. S3 in Supplement shows an example in which a post-operation incubation of a walker-track sample restored its fluorescence to the pre-operation level. Hence, the plus-end buildup is a truly non-equilibrium effect due to the walker's physical action.

The fluorescence data in Figs. 3, 4 were collected immediately before and after each individual UV irradiation; the pattern of successive drop indicates that the walker transforms the UV-induced leg dissociation primarily into a directional walking on the track rather than brutal derailment. After a leg is dissociated by UV, the walker can be derailed via subsequent dissociation of the other leg from D1-D1* duplex by thermal fluctuation (see Fig. 1E, panels 2, 6) or from D2-D2* by another UV irradiation (panel 3). Either channel of derailment occurring at the plus end will cause a fluorescence rise immediately after UV irradiation. The observed opposite pattern rules out derailment as a major process (see Fig. S2 for more data). Furthermore, derailed walkers lose direction; their subsequent re-binding to the identical sites produces no preferential plus-end accumulation. Derailment is suppressed by binding of the dissociated leg (e.g. from panel 3 to panel 4 or 1, Fig. 1E), and by the 25 $^o$C operation temperature that is below the melting temperature of the *cis*-form D2-D2* duplex (~ 32 $^o$C)[24] to enable partial D2-D2* hybridization even under UV. This suppresses derailment from the single-leg D1-D1* binding (panels 2, 6) though the desired leg dissociation becomes slower too (panel 5).

The operation-induced signal drop of ~ 54% in Fig. 4 provides evidence for the walker's full-step translocation from one composite site to another towards the plus end. In a sample of walker plus the two-site track, the loop state can occur at either site but only one inter-site bridge state can exist (B1 or occasionally B3). Since the loop is lower than B1 in free energy, and also lower than B3, the population ratio of the bridge state over either loop state in the pre-operation equilibrated sample is no more than 1 by Boltzmann's law. The bridge thus accounts for less than 1/3 of the walker population prior to operation. Translocation of the entire bridge population to the loop at the plus end would add less than 1/6 of walker legs to the quencher's vicinity. This contributes, at most, a fluorescence drop of 1/6 of the total emission of the walkers. Since the pre-operation fluorescence is ~ 51% of the total emission, the entire bridge translocation would produce a signal drop of no more than 33%. The remaining ~ 21% drop must be due to translocation of the loop population from one composite site to the other at the plus end.

In summary, this study demonstrates a design principle by which a symmetric bipedal walker on a track of identical composite binding sites gains a direction by matching the walker's size against the inter-site spacing. Conceptually, this amounts to a previously proposed mechanical breaking[19] of symmetry. The symmetry breaking facilitates a selective rear leg dissociation that serves as a nano-ratchet to rectify directional motion. Notably, the ratchet for inter-site stepping is amplified from a local, intra-site asymmetry that is merely one fifth of the inter-site spacing and made by a minimum of two binding components. Derived from biology and based on mechanical effects, the design principle is adaptable for use in other nanomachines. Besides, the walker is advantageous for biomedical applications as it is free of chemical wastes, remotely controlled by light, and requires a low-level UV irradiation (Supplement).

**Figure captions**

**Figure 1. Design principle of the walker.** The dashed circle in **B** shows a composite binding site formed by a pair of nearest D1* and D2* footholds. The star (*) marks a complementary sequence. Panels 4-8 in **E** show how the walker obtains a direction for inter-site walking under a light operation that alternately disrupts and restores D2-D2* duplex. Panels 1-3 show how an intra-site loop state responds to the operation.

**Figure 2. Assembly of the walker-track**. Gel images obtained using native PAGE for purified walker, tracks of different lengths as indicated, and walker-track binding complexes.

**Figure 3. Fluorescence detection of the walker in operation**. The three panels show experiments that use the same amount of walker sample in an equimolar walker-track mix but for different tracks as indicated. The data for zero time are from the equilibrated sample before the operation. Seven rounds of UV irradiation were applied, which lasted 10, 20, 30, 30, 30, 30, and 30 minutes, each followed by 20 seconds of visible irradiation during which the fluorescence was collected.

**Figure 4. Fluorescence of the walker under a 10-hours operation on two-site tracks**. The light operation is the same as for Fig. 3 except that the 20 rounds of UV irradiation each lasted 30 minutes. The shown signals were obtained by integrating the fluorescence spectra over the wavelength (433 nm – 627 nm). A control experiment for walkers alone under the same operation yielded a fluorescence change within 1%.

ACKNOWLEDGMENT

We thank Nadrian C. Seeman for providing SEQUIN code that assisted the early stage of this study, and thank Jie Yan, Tianhu Li, Thorsten Wohland, Matthew Lang, Daniel Lubrich for valuable discussions or comments. This work is partially supported by FRC grants under R-144-000-244-133 and R-144-000-259-112 (to ZSW).

**Figure 1**

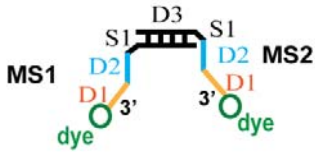
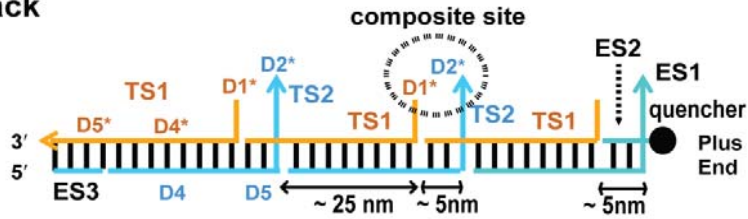
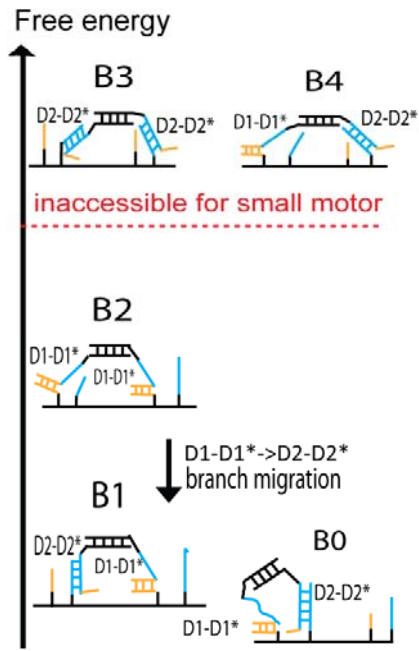
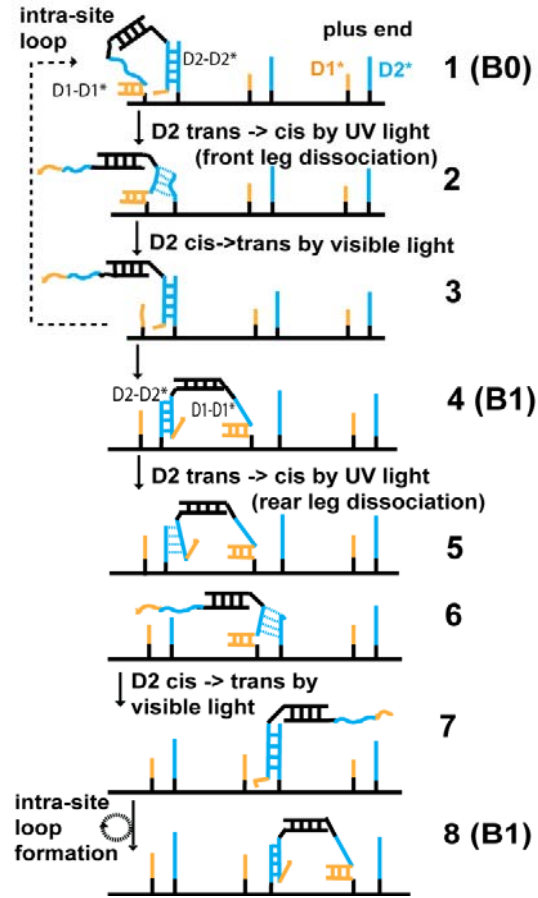

**Figure 2**

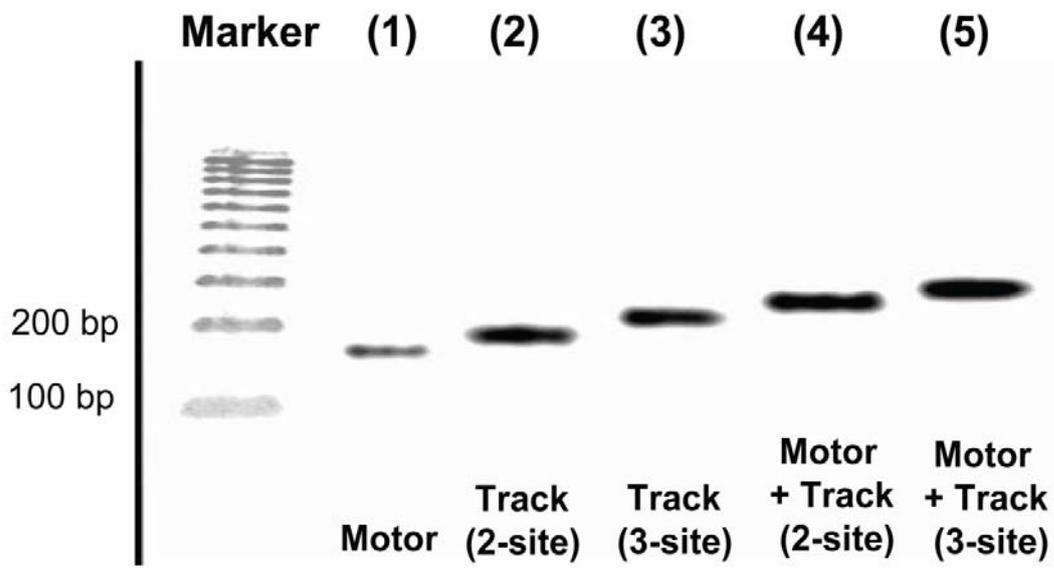

**Figure 3**

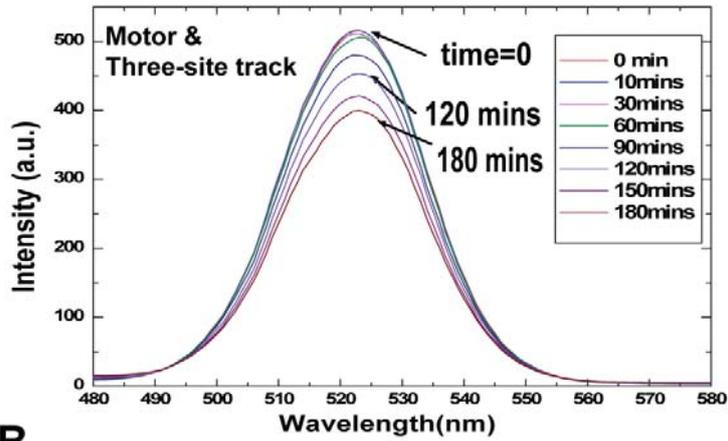

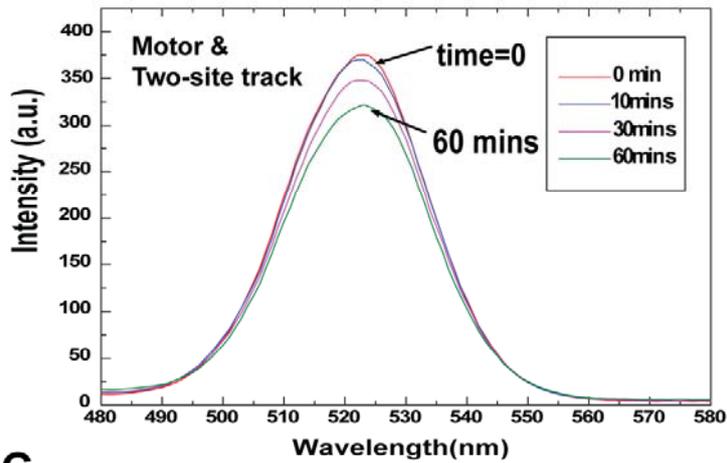

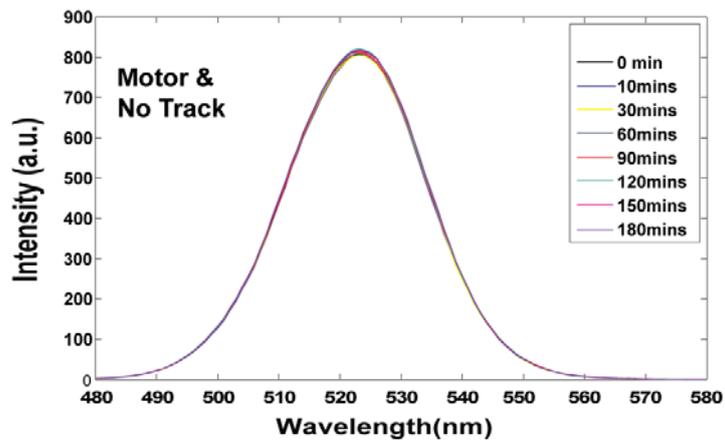

**Figure 4**

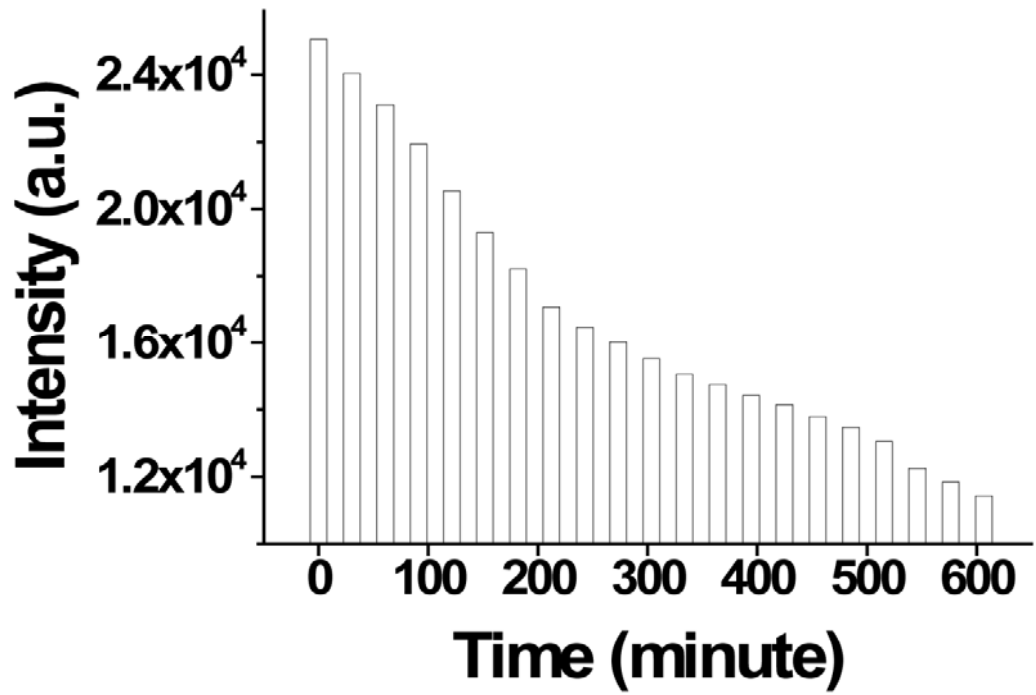